\documentclass[onecolumn,pra, superscriptaddress,longbibliography,margin=1mm 12pt]{revtex4-2} 

\usepackage{graphicx}
\usepackage{bm}
\usepackage{amsmath}
\usepackage{environ}
\usepackage{appendix}
\usepackage{physics}

\usepackage[nopar]{lipsum}

\usepackage{amsthm}
\usepackage{amssymb}
\usepackage{amsfonts}
\usepackage{fancyhdr}
\usepackage{slashed}
\usepackage{bbm}
\usepackage{latexsym,epsfig,bbm}
\usepackage[colorlinks=true,urlcolor=blue,citecolor=blue]{hyperref}

\usepackage{color}
\definecolor{blue}{rgb}{0,0.2,1}

\definecolor{red}{rgb}{0.9,0,0}

\newcommand{\vect}[1]{\boldsymbol{#1}}

\begin{document}


\title{Quantum simulation of discrete linear dynamical systems and simple iterative methods in linear algebra via Schr\"odingerisation}

\date{\today}

\author{Shi Jin}
\affiliation{Institute of Natural Sciences, Shanghai Jiao Tong University, Shanghai 200240, China}
\affiliation{School of Mathematical Sciences, Shanghai Jiao Tong University, Shanghai 200240, China}
\affiliation{Ministry of Education Key Laboratory in Scientific and Engineering Computing, Shanghai Jiao Tong University, Shanghai 200240, China}
\affiliation{Shanghai Artificial Intelligence Laboratory, Shanghai, China}

\author{Nana Liu}
\email{nana.liu@quantumlah.org}
\affiliation{Institute of Natural Sciences, Shanghai Jiao Tong University, Shanghai 200240, China}
\affiliation{Ministry of Education Key Laboratory in Scientific and Engineering Computing, Shanghai Jiao Tong University, Shanghai 200240, China}
\affiliation{University of Michigan-Shanghai Jiao Tong University Joint Institute, Shanghai 200240, China.}
\affiliation{Shanghai Artificial Intelligence Laboratory, Shanghai, China}


\begin{abstract} 
Quantum simulation is known to be capable of  simulating certain dynamical systems in continuous time -- Schr\"odinger's equations being the most direct and well-known -- more efficiently than classical simulation. Any linear dynamical system can in fact be transformed into a system of Schr\"odinger's equations via a method called Schr\"odingerisation \cite{schr2}. Building on the observation that iterative methods in linear algebra, and more generally discrete linear dynamical systems, can be viewed as discrete time approximations of dynamical systems which evolve continuously in time, we can apply the Schr\"odingerisation technique. Thus quantum simulation can be directly applied to the continuous-time limits of some of the simplest iterative methods. This applies to general (explicit) iterative schemes or discrete linear dynamical systems. In particular, we introduce the quantum Jacobi and quantum power methods for solving the quantum linear systems of equations and for estimating the maximum eigenvector and eigenvalue of a matrix respectively. The proposed quantum simulation can be performed on either discrete-variable quantum systems or on hybrid continuous-variable and discrete-variable quantum systems. This framework provides an interesting alternative method to solve linear algebra problems using quantum simulation. 
\end{abstract}
\maketitle 

\section{Introduction}
Quantum simulation for quantum dynamics evolving in continuous time is one of the earliest proposed tasks for quantum devices \cite{feynman2018simulating, daley2022practical}. In addition to the simulation of quantum dynamics, quantum devices are also potentially important for solving linear algebra problems, including the linear systems of equations problem \cite{harrow2009quantum, clader2013preconditioned, childs2017quantum,gilyen2019quantum, subacsi2019quantum} and identifying maximum eigenvectors and eigenvalues \cite{lloyd2014quantum, nghiem2022quantum}. In fact, the quantum analogue of the linear systems of equations problem is known to be BQP-complete \cite{harrow2009quantum} -- the hardest task for a quantum computer. However, these problems, which are discrete in nature, are not typically viewed under the lens of continuous-time quantum dynamics. Thus, an intriguing question is whether and how problems in linear algebra can be viewed more directly in terms of the continuous-time evolution of dynamical quantum systems. These quantum systems might then be viewed as alternative building blocks to quantum computation. 

Transforming problems of a discrete nature into dynamical problems evolving in continuous time can often be a useful tool. For instance, combinatorial search problems can be turned into optimisation problems that are solved in continuous time through quantum adiabatic computation \cite{farhi2000quantum, aharonov2008adiabatic} Here one begins with the known ground state of one Hamiltonian which adiabatically evolves to the ground state of another Hamiltonian whose ground state is the solution to the original problem. It can also be a starting point for solving problems in linear systems of equations \cite{subacsi2019quantum}.

We take an alternative route to the continuous-time perspective by considering very simple iterative methods in linear algebra. Iterative methods, and more generally discrete linear dynamical systems \cite{KM2002discrete, galor2007discrete}, evolve by discrete time steps. However, one can transform the iteration process into dynamical systems by taking the continuous-time limit. This approach has been adapted for instance in relating gradient flow to gradient descent algorithms. In our recent work \cite{schr2} it was shown how a general dynamical system can be mapped directly into quantum dynamical systems via a method called Schr\"odingerisation -- i.e., turning a non-Schr\"odinger equation into systems of Schr\"odinger's equations. As a result, the continuous limit of the iterative procedure can thus be subsequently simulated on a quantum simulator.

Our formalism allows potential implementations on both discrete and hybrid continuous-variable discrete-variable (CV-DV) quantum systems \cite{andersen2015hybrid}. While most proposals for quantum computation is fully discrete in the sense that it operates only on qubits, the underlying physics of these systems are in fact continuous-variable in nature. For example, quantum harmonic oscillators are not qubits but quantum modes or qumodes. It is then natural to take advantage of their continuous nature and hybrid continuous-variable and discrete-variable systems can provide an alternative path to quantum computation \cite{andersen2015hybrid, lloyd2003hybrid, liu2016power}. 

We start from general (explicit) iterative methods, or more generally discrete linear dynamical systems, which will be viewed as solutions of a system of homogeneous linear differential equations. These can be subsequently solved by quantum simulation following the Schr\"odingerisation methods proposed in \cite{schr2}. The long time behavior, which corresponds to the convergence of the iterative method, gives the steady state of the continuous system, which corresponds to the  limit of the iterations. The quantum simulation procedure will be presented separately for a Hermitian and a non-Hermitian operator and the cost of this approach will be analyzed. We then apply this formalism to the Jacobi method and the power method in computational linear algebra.

The Jacobi method is one of the simplest iterative methods for solving a linear system of equations \cite{GvL}. It is a stationary iterative algorithm whose steady state solution solves the linear systems of equations problem. A costly step in the procedure is the matrix multiplication at each iterative step. We show how continuous-time quantum simulation can be applied to the approximation of the steady state solution and to solve the quantum analogue of the linear system of equations problem -- preparing quantum states whose amplitudes are proportional to the classical solutions. Since quantum simulation is used to replace the classical matrix multiplication at each iterative step, the Jacobi method can be made more efficient on a quantum simulator. We call this the {\it quantum Jacobi method}. Given the quantum state solution, some of the classical solutions to the original problem can still be efficiently extracted. This procedure can in principle also be generalised to other iterative methods. 

Another important problem in linear algebra is identifying the maximum eigenvalue and the corresponding eigenvector of a given matrix. A very simple and effective iterative method  is the power method. Similarly to the Jacobi method, a costly step is the matrix multiplication at each iterative step, which we replace with the continuous-time quantum simulation. Here our quantum simulation algorithm prepares the quantum state whose amplitudes form a vector that is the eigenvector corresponding to the maximum eigenvalue. This protocol will be called  the {\it quantum power method}. The resulting quantum state can then be used to extract the maximum eigenvalue efficiently.

In the last section, we discuss comparisons with some other methods.


\section{Quantum iterative methods for discrete linear dynamical systems} \label{sec:generaliteration} 

The execution of any iterative method in linear algebra (or a discrete linear dynamical system \cite{galor2007discrete}) involves the evolution of a state $y$ for a long enough (discrete) time so the state approaches the steady state, i.e., the iteration converges.  General iterative methods for linear problems or linear dynamical systems take the form
\begin{align} \label{eq:generaly}
    y_{k+1}=G y_k +g, \qquad k \in \mathbb{Z}^{+} \cup \{0\}
\end{align}
where $y_k, g \in \mathbb{R}^d$ and $G$ is a $d \times d$ matrix. Here $k$ labels the time step in the iterative approach. To simplify the form, we can define an augmented vector $x_k=(y_k, 1)^T$ which evolves a homogeneous system according to 
\begin{align}
    x_{k+1}=C x_k, \qquad C=\begin{pmatrix}
        G & g \\
        \vect{0}^T & 1
    \end{pmatrix}
\end{align}
where $C$ is a $(d+1)\times (d+1)$ matrix and $\vect{0}^T=(0,...,0)$ is $d$-dimensional. We can rewrite this in the form
\begin{align} \label{eq:xdynamics}
x_{k+1}-x_k=\begin{pmatrix} 
y_{k+1}-y_k\\
0
\end{pmatrix}=(C-I)x_k.
\end{align}
From this form it is simple to see, for instance, that the iterative method converges or reaches its steady-state when $y_{k+1} \approx y_k$, i.e., $ y_{k+1}-y_k=(G-I)y_k+g \rightarrow 0$, which coincides with identifying the ground state of $C-I$.

Since $x_{k+1}-x_k=(C-I)x_k=(C-I)C^k x_0=C^k(Cx_0-x_0)=C^k(x_1-x_0)$, then $\|x_{k+1}-x_k\|\leq \|C\|^k\|x_1-x_0\|$. When the spectral radius of $C$ is $r(C)<1$, then $\|C\|<1$ for any subordinate norm $\|\cdot\|$, so the convergence rate $\|C\|^k$ for state $x$ is exponential with time-step $k$ \cite{GvL}. We can convert Eq.~\eqref{eq:xdynamics} into a dynamical equation by converting $k$ into a continuous time $t \in \mathbb{R}^+ \cup \{0\}$, so the iterative relation $x_{k+1}-x_k=(C-I)x_k$ is transformed into a corresponding system of linear ordinary differential equations (ODEs)
\begin{align} \label{eq:odebasic}
    \frac{dx}{dt}=(C-I)x, \qquad x(t=0)=x_0.
\end{align}
Note that since $r(C)<1$, then $C-I$ has negative eigenvalues. This means the corresponding ODE system is contractive, thus the initial value problem is stable and the solution decays to the steady state exponentially in $t$. 

To exploit quantum simulation, our aim is to instead prepare the quantum state $|x\rangle=(1/\|x\|)\sum_{i=0}^{d}x_i|i\rangle$ where $\{x_i\}_{i=0}^d$ are the entries of the vector $x$, $\|x\|$ is its $l_2$-norm and $\{|i\rangle\}_{i=0}^d$ is an orthonormal basis. Here the classical solution $x(t)$ to our continuous-time iterative procedure is encoded in the amplitude of the quantum state. Since the iterative procedure is now a dynamical system, we can simulate its solution $|x\rangle \propto x$ by dynamical methods that is continuous in $t$. In particular, we can simulate Eq.~\eqref{eq:odebasic} directly with quantum simulation using the Schr\"odingerisation technique (i.e., turning a non-Schr\"odinger equation into a system of Schr\"odinger's equations) we recently introduced in \cite{schr2}.

Below we summarise how Schr\"odingerisation can be used to simulate the evolution of $x$ for both Hermitian and non-Hermitian $C$. This allows us to apply quantum simulation to directly simulate the evolution of $x$ -- and hence $y$ -- which can be used to estimate the solutions of the quantum linear systems of equation problem and to identify the maximum  eigenvalue and the corresponding eigenvector  of a given matrix.

\subsection{Evolution of $x$ for Hermitian $C$} \label{sec:A}

We begin by considering the evolution equation of the form 
\begin{align} \label{eq:odeH}
    \frac{dx}{dt}=-H x, \qquad x(0)=x_0,
\end{align}
where $H=-(C-I)=H^{\dagger}$, which implies the scenario $G=G^{\dagger}$ and $g=0$ in Eq.~\eqref{eq:generaly}. To use quantum simulation for Eq.~\eqref{eq:odeH}, we can use the Schr\"odingerisation technique \cite{schr2}, which introduces a single extra dimension $p$, and define $w=\exp(-p)x$ for $p>0$. This is referred to as the warped phase transformation. One then solves for $w$ for all $p \in (-\infty, \infty)$ that satisfies
\begin{align}\label{phase-eqn}
\begin{cases}
    &\frac{\partial w}{dt}=H\frac{\partial w}{dp} \\
    & w(0)=e^{-|p|}u_0.
    \end{cases}
\end{align}
We point out that this system is hyperbolic --hence the initial value problem \eqref{phase-eqn} is well-posed -- if $H$ is diagonalizable in real space, namely it has only real eigenvalues and a complete set of eigenvectors \cite{Lax}. Moreover, if we assume $H$ is positive definite, then we do not need a boundary condition for $H$ in \eqref{phase-eqn} at $p=0$. 

By taking the  Fourier transform of $w$ with respect to $p$, denoted $\tilde{w}$, one arrives at a system of Schr\"odinger's equations
\begin{align} \label{eq:wtilde}
\begin{cases}
   & i\frac{d\tilde{w}}{dt}=\eta H\tilde{w}, \qquad \forall \eta \in (-\infty, \infty) \\
   & \tilde{w}(0)=(1/(\pi(1+\eta^2)))x_0.
   \end{cases}
\end{align}
For instance, we can choose to consider hybrid discrete variable-continuous variable (CV-DV) states $|\tilde{w}(0)\rangle_{CV-DV}=(1/\|f(\eta)\|)\int_{\-\infty}^{\infty} d \eta f(\eta)|\eta\rangle |x_0\rangle$, $f(\eta) \propto 1/(1+\eta^2)$, $|x_0\rangle=(1/\|x_0\|)\sum_{i=0}^{D-1}x_{0, i}|i\rangle$ and $\| \cdot \|$ is the $l2$-norm of $(\cdot)$. Here the state $(1/\|f(\eta)\|)\int_{\-\infty}^{\infty} d \eta f(\eta)|\eta\rangle$ is a continuous-variable quantum state because it is an infinite dimensional state, since $\eta$ is a continuous degree of freedom. This is known as a quantum mode or qumode. Qumodes can also be used as information carriers and is an alternative to using qubits \cite{braunstein2005quantum, adesso2014continuous}. For example, we can let $\eta$ be represented by position eigenstates $|q\rangle$ of the quantum harmonic oscillator with position operator $\hat{q}$. The momentum operator $\hat{p}$ has corresponding eigenstates $|p\rangle$ where $[\hat{q}, \hat{p}]=i$ and $|p\rangle$ is the continuous-variable quantum Fourier transform of $|q\rangle$. This means that the amplitudes of $|p\rangle$ are the continuous Fourier transform of amplitudes of $|q\rangle$. Continuous-variable quantum Fourier transforms, unlike discrete quantum Fourier transforms, are particularly straightforward to implement for instance on quantum optical systems since it is just a change from the position basis to the momentum basis. On the other hand, $|x_0\rangle$ is finite $D$-dimensional quantum state, realised by a discrete $\log_2(D)$ number of qubits. The evolution of $|\tilde{w}(t)\rangle_{CV-DV}=\exp(-iH \otimes \hat{q}t)|\tilde{w}(0)\rangle_{CV-DV}$ is thus governed by a hybrid CV-DV quantum gate $\exp(-iH_{CV-DV}t)$ with $H_{CV-DV}=H \otimes \hat{q}$. These quantum gates can be considered as an alternative building block to quantum computation which might be realisable in quantum optics, trapped ion systems and superconducting circuits \cite{andersen2015hybrid, van2008hybrid, sutherland2021universal, lloyd2003hybrid, liu2016power}. These gates can also be used to implement alternative quantum realisations of DQC1, quantum phase estimation, Shor's factoring algorithm \cite{liu2016power} and sensing \cite{elliott2017continuous}.

We can also perform quantum simulation on purely discrete quantum systems, qubits, by discretising $\eta=\eta_l$ in increments of size $\Delta \eta$ and $l=-N/2+1,...,N/2$. The evolution of the quantum state $|\tilde{w}(t)\rangle$ then proceeds according to the Schr\"odinger equation
\begin{align}
    \begin{cases} \label{eq:wtildediscrete}
       &  i \frac{d}{dt}|\tilde{w}(t)\rangle=(H \otimes D)|\tilde{w}(t)\rangle \\
       & |\tilde{w}(0)\rangle=\frac{\|x_0\|}{\|\tilde{w}(0)\|}\sum_{j=-N/2+1}^{N/2} (1/(\pi(1+\eta_j^2)))|j\rangle |x_0\rangle
    \end{cases}
\end{align}
where $|\tilde{w}(t)\rangle=(1/\|\tilde{w}(t)\|)\sum_{i=0}^{D-1}\sum_{l=-N/2+1}^{N/2} \tilde{w}_i(t, \eta_l)|i\rangle|l\rangle$ and $D=\text{diag}(-N/2+1,..., N/2)$. 

After preparing $|\tilde{w}(0)\rangle$, one needs to perform quantum simulation for the evolution $\exp(-i(H \otimes D)t)$ to obtain $|\tilde{w}(t)\rangle$. Then an inverse quantum Fourier transform with respect to $\eta$ is performed  to obtain state $|w(t)\rangle=(1/\|w(t)\|)\sum_{i=0}^{D-1}\sum_{l=-N/2+1}^{N/2} w_i(t,p_l)|i\rangle |l\rangle$. We can then project onto the $p>0$ states, where $p=p_k$ is also discretised with $k=-N/2+1,...,N/2$. We note that projection onto a single $p=p^*$ for some $p^*>0$ is an alternative approach that also gives the same final outcome $|x(t)\rangle$. Using $\hat{P}=I \otimes \sum_{l=0}^{N/2}|l\rangle \langle l|$, where $\hat{P}|w(t)\rangle=(\|x(t)\|\|\exp(-p)\|/\|w(t)\|)|\exp(-p)\rangle |x(t)\rangle$ where $|\exp(-p)\rangle=(1/\|\exp(-p)\|)\sum_{l=0}^{N/2} \exp(-p_l)|l\rangle$ and $\|\exp(-p)\|=\sum_{l=0}^{N/2} \exp(-2p_l)$. Thus one can retrieve $|x(t)\rangle$ from $|w(t)\rangle$ with probability $(\|x(t)\|\|\exp(-p)\|/\|w(t)\|)^2 \sim (\|x(t)\|/\|x_0\|)^2$, giving a cost where this probability can also be quadratically boosted \cite{schr2}. 

We can use the above quantum simulation protocol to prepare the approximation $|x(t=t_f)\rangle$ of the steady state solution of $x$ (and thus of $y$) to a high precision. This would be the same length of time $t=t_f$ used in the quantum simulation problem in Eq.~\eqref{eq:wtilde} and Eq.~\eqref{eq:wtildediscrete}. To estimate this $t_f$, we first expand the initial state $|x_0\rangle=(1/\|x_0\|)\sum_{i=0}^{D-1} (x_0)_i |i\rangle$ as 
$|x_0\rangle=\frac{1}{\|x_0\|}\sum_{j=0}^{D-1}\alpha_j|E_j\rangle, \alpha_j \in \mathbb{C}$, 
with normalisation defined as $\| \cdot \|^2=\sum_{i=0}^{D-1} |(\cdot)_i|^2$. Since $H$ is Hermitian, it has orthonormal eigenvectors $\{|E_j\rangle\}_{j=0}^{D-1}$ with corresponding eigenvalues $\{E_j\}_{j=0}^{D-1}$, where $E_0<...<E_{D-1}$.  Then 
\begin{align}
    |x(t)\rangle=\frac{\|x_0\|}{\|x(t)\|}e^{-Ht}|x_0\rangle=\frac{1}{\|x(t)\|}\sum_{j=0}^{D-1}\alpha_j e^{-E_j t}|E_j\rangle.
\end{align}
We can write 
\begin{align} \label{eq:unorm}
    \|x(t)\|^2=\|e^{-Ht}u_0\|^2=\left\|\sum_{j=0}^{D-1} \alpha_j e^{-E_j t}|E_j\rangle\right\|^2=\sum_{j=0}^{D-1} |\alpha_j|^2e^{-2E_j t}=e^{-2E_0t}(|\alpha_0|^2+|\alpha_1|^2e^{-2 t \Delta}+L)
\end{align}
where the spectral gap is $\Delta_C=E_1-E_0>0$ and $L=\sum_{k=2}^{D-1} |\alpha_k|^2\exp(-2 t (E_k-E_0))$. To determine how long it takes to evolve such a quantum system, we say $t=t_f$ when the fidelity between $|x(t)\rangle$ and the true ground state $|x_g\rangle=|E_0\rangle$ of $H$ is greater or equal to $1-\delta$, $\delta>0$, i.e.,
\begin{align} \label{eq:fidelitycondition}
   \mathcal{F}(|E_0\rangle, |x(t_f)\rangle)= |\langle x(t_f)|E_0\rangle|^2=\frac{|\alpha_0|^2e^{-2E_0 t_f}}{\|x(t_f)\|^2}\geq 1-\delta.
\end{align}
Combining Eqs.~\eqref{eq:fidelitycondition} and ~\eqref{eq:unorm}, since $\delta \ll 1$, we get the condition
\begin{align}
    t_f\geq \frac{1}{2\Delta_C}\ln \left(\frac{|\alpha_1|^2(1-\delta)}{|\alpha_0|^2 \delta }\left(\frac{1}{1-L(1-\delta)/(|\alpha_0|^2\delta)}\right) \right).
\end{align}
When $L$ is  small (relative to $|\alpha_0|^2$), we mean $L \ll \delta |\alpha_0|^2/(1-\delta)$ and $\delta \ll 1$, we have $t_f \gtrsim (1/(2\Delta))\ln(|\alpha_1|^2/(\delta |\alpha_0|^2))$. For larger $L$ we simply include more terms in $L$ to find $t_f$. If the first two eigenstates $|E_0\rangle$ and $|E_1\rangle$ dominate so that $|\alpha_1|^2 \approx 1-|\alpha_0|^2$ then 
\begin{align}\label{heat-tfinal}
    t_f \gtrsim \frac{1}{2\Delta_C} \ln \left(\frac{1}{\delta} \left(\frac{1}{|\alpha_0|^2}-1\right)\right).
\end{align}

The total cost in this ground state $|x_g\rangle$ estimation is the cost (a) in preparing the initial state $|\tilde{w}(0)\rangle$ in Eq.~\eqref{eq:wtildediscrete}, (b) in the quantum simulation for the system Eq.~\eqref{eq:wtildediscrete}, and (c) the cost in the projection back to $|u(t_f)\rangle$ from $|\tilde{w}(t_f)\rangle$. Here we choose to consider the case where $\eta$ is discretised. 

Preparing the initial state $|\tilde{w}(0)\rangle$ in Eq.~\eqref{eq:wtildediscrete} requires the preparation of $|u_0\rangle$ and the state $\propto \sum_{j=-N/2+1}^{N/2}(1/(\pi(1+\eta_j^2)))|j\rangle$, which is the quantum Fourier transform of the state $|w(0)\rangle \propto \sum_{l=-N/2+1}^{N/2}\exp(-|p_l|)|l\rangle$. Although neither of these are sparse states -- approximation into sparse states would yield larger errors than desired -- and order $O(N)$ resources may be needed, this resource state is the same for every problem, unlike $|u_0\rangle$. Thus it is reasonable to assume this state as being given while $|u_0\rangle$ is prepared. A simple choice could be $|u_0\rangle=|1\rangle \otimes |x_0\rangle$. If the number of non-zero entries in $|u_0\rangle$ is for instance $\log N$, then the preparation is efficient \cite{zhang2022quantum}. As we will see later, only a single copy of this initial state is necessary. 

The cost in the quantum simulation step would differ depending on the simulation method used. For instance, by simulating $\exp(-i(H \otimes D)t)$ with digital quantum simulation where the unitary is decomposed into one and two-qubit gates, this can carry a complexity cost (e.g. \cite{berry2015hamiltonian}) $\tilde{\mathcal{O}}(s_H \|H \otimes D\|_{max}t)=\tilde{\mathcal{O}}(s_C \| C-I\|_{max} N \otimes \|t)=\tilde{\mathcal{O}}(s_C \|C\|_{max} t_f/\epsilon)$, which for $t=t_f$ gives a cost $\tilde{\mathcal{O}}(s_C \|C_I\|_{max}/(\epsilon\Delta_H))$ where $s_{(\cdot)}$ is the sparsity of $(\cdot)$, $\| \cdot \|_{max}$ denotes its max-norm and $\epsilon \sim 1/N$ is the error in the solution coming from the discretisation of $\eta$. Here $\Delta_C$ is the spectral gap of $C$.

In analogue quantum simulation it might be possible to create a quantum system that naturally realises the CV-DV gate $\exp(-i(H \otimes \hat{q})t)$. However, the minimum time cost is still of order $t_f=\tilde{\mathcal{O}}(1/\Delta_C)$.  

To retrieve $|x(t_f)\rangle$ we require an inverse quantum Fourier transform on $|\tilde{w}(t)\rangle$ before projecting onto either the positive $p$ values or a particular positive $p=p^*$. Alternatively, we can instead perform a rotation and use amplitude amplification and requires a complexity cost $\|x(0)\|/\|x(t_f)\|$ in obtaining $|x(t_f)\rangle$. This means only a single copy of the state $|\tilde{w}(t_f)\rangle$ is required. For instance, if we use assume $\|x(0)\|=1$, $E_0 \leq 0$ and $\delta \ll 1$, then from Eq.~\eqref{eq:fidelitycondition} we see $\|x(0)\|/\|x(t_f)\|=\mathcal{O}(1/|\alpha_0|)$. 

We note that if our purpose is not to prepare the ground state itself but only to recover an expectation value of $|x(t_f)\rangle$ with respect to some observable $O$, then it is not necessary to convert $|\tilde{w}(t)\rangle \rightarrow |w(t)\rangle$ by a discrete quantum Fourier transform, since we can extract $\langle \tilde{w}(t, \eta)|(I \otimes O)|\tilde{w}(t, \eta)\rangle=\langle w(t, p)|(I \otimes O)| w(t,p)\rangle \propto \langle x(t)|O|x(t)\rangle$. 

In the more general case where $C \neq C^{\dagger}$, (i.e. $G \neq G^{\dagger}$ and $g \neq 0$ in general), it is possible to perform a dilation by defining the $2(d+1) \times 2(d+1)$ Hermitian matrix $H$ as $H=\begin{pmatrix}
        0 & -(C-I) \\
        -(C^{\dagger}-I) & 0
    \end{pmatrix}$
and we then solve for $du/dt=-Hu$ where $u$ is a $2(d+1)$-dimensional vector. In this case, $u(t)$ does not map onto $x(t)$ in a simple way for general $t$, except in the steady state. We note that $C-I$ has a non-degenerate ground state $x_g$ for the iterative method to converge to a unique answer. We can denote the ground state/s of $C^{\dagger}-I$ as $x'_g$, which may or may not be degenerate in general. However, the ground states $u_g$ of $H$ are degenerate and belong to three different classes: $u_g=(0, x_g)^T$, $u_g=(x'_g, x_g)^T$ and $u_g=(x'_g, 0)^T$. If $u_g=(0, x_g)^T$ or $u_g=(x'_g, x_g)^T$, then $x_g$ is straightforward to retrieve and the previous protocol can be applied. However, if the ground state $u_g=(x'_g, 0)^T$, then this does not allow us to retrieve $x_g$. This means that this dilation method for preparing the approximate ground state of $C$ for $C \neq C^{\dagger}$ is only applicable under the assumption we do not reach the latter ground state, given suitable initialisation for $u_0=u(t=0)$. For more general cases of $C \neq C^{\dagger}$ which may not satisfy such assumptions, we use an alternative method which we outline in the next section.

\subsection{Evolution of $x$ for non-Hermitian $C$} \label{sec:B}

In the more general case where $C\neq C^{\dagger}$, which occurs when $G \neq G^{\dagger}$ and/or $b \neq 0$, we can apply the Schr\"odingerisation approach \cite{schr2} directly onto Eq.~\eqref{eq:odebasic}. Since the solution to Eq.~\eqref{eq:odebasic} is $x(t)=\exp((C-I)t)x_0$, we use a quantum simulation procedure to prepare the quantum state
\begin{align}   \label{eq:xtc}
|x(t)\rangle=\frac{\|x_0\|}{\|x(t)\|}e^{(C-I)t}|x_0\rangle. 
\end{align}
Here one can always decompose $C-I=C_1+iC_2$ as a sum of a Hermitian matrix $C_1=(1/2)(C+C^{\dagger}-2I)=C_1^{\dagger}$ and an anti-Hermitian matrix $iC_2$, where $C_2=(1/2i)(C-C^{\dagger})=C_2^{\dagger}$ is itself Hermitian. In the special case where $C_1, C_2$ commute, i.e., $[C_1, C_2]=0$, then $|x(t)\rangle=\exp(iC_2t)(\|x_0\|/\|x(t)\|)\exp(C_1)|x_0\rangle$. 
Since $C_1^{\dagger}=C_1$, we can apply the method in the previous section with $H=-C_1$. After this step, quantum simulation can be used to realise the unitary operation $\exp(iC_2t)$ to obtain $|x(t)\rangle$. In other cases where the commutator $[C_1,C_2]$ takes simple forms, e.g. $\propto I$, Campbell-Baker-Hausdorff relations can be evoked so the method in Section~\ref{sec:A} can still be applied beyond preparing the steady state solution.  

In the most general cases $[C_1, C_2] \neq 0$, we must proceed differently by directly Schr\"odingerising the evolution $x(t)=\exp((C-I)t)x_0$. Here we define the vector $v(t,p)=\exp(-p)x(t)$ for $p>0$. For $p \in (-\infty, \infty)$, $v$ satisfies
\begin{align} \label{eq:vequation}
    \begin{cases}
        \frac{\partial v}{dt}=-C_1\frac{\partial v}{dp}+iC_2 v\\
        v(0)=e^{-|p|}x_0.
    \end{cases}
\end{align}
Here, like in the preceding section, we assume $C_1$ to be diagonalizable in real space, and negative definite. 

Defining $\tilde{v}(t,\eta)=\mathcal{F}(v(t,p))$ as the Fourier transform of $v$ with respect to $p$, where $\eta \in \mathbb{R}$ is the Fourier mode of $p$, then
\begin{align}
\begin{cases}
    i\frac{d \tilde{v}}{dt}=-(\eta C_1+C_2)\tilde{v}, \qquad \forall \eta \in (-\infty, \infty) \\
    \tilde{v}(0)=\frac{1}{\pi(1+\eta^2)}x_0
    \end{cases}.
\end{align}
This is now a system of Schr\"odinger's equations, one for each $\eta$, with the corresponding Hamiltonian $-(\eta C_1+C_2)$.  This can be implemented with a hybrid CV-DV Hamiltonian $H_{CV-DV}=-(C_1 \otimes \hat{q}+C_2 \otimes I)$. This reduces to Eq.~\eqref{eq:wtilde} when $C=C^{\dagger}$, i.e., $C_2=0$.

Like in the previous section, we can also transform this completely into a discrete systems, where we use discrete Fourier transform with discrete  $\eta$ to obtain the following evolution 
\begin{align}
\begin{cases}
i\frac{d |\tilde{v}(t)\rangle}{dt}=-(C_1 \otimes D+C_2 \otimes I)|\tilde{v}(t) \rangle=H_{tot}|\tilde{v}(t) \rangle \\
|\tilde{v}(0)\rangle=\frac{\|x_0\|}{\|\tilde{v}(0)\|}\sum_{j=-N/2+1}^{N/2} (1/(\pi(1+\eta^2_j))|x_0\rangle
\end{cases}
\end{align}
where $|\tilde{v}(t)\rangle=(1/\|\tilde{v}(t)\|) \sum_{i=0}^d \sum_{l=-N/2+1}^{N/2} \tilde{v}_i(t, \eta_l)|i\rangle |l\rangle$ and 
\begin{align} \label{eq:Htot}
    H_{tot}=-C \otimes (D-iI)/2-C^{\dagger} \otimes (D+iI)/2+I\otimes D=H^{\dagger}_{tot}.
\end{align}
Similarly to the previous section, from $|\tilde{v}(t)\rangle$, the state $|x(t)\rangle$ can be recovered by an inverse (discrete) Fourier transform and a projection onto positive $p$ states or a single $p=p^*$. 

Since $x(t)=(y(t), 1)^T$, the state $|y(t)\rangle$ is easily obtained from $|x(t)\rangle$. This method holds for any $t$, not only for very large $t$ where one approaches the steady state of $C-I$. 

 The cost in the digital quantum simulation to prepare $|\tilde{v}(0)\rangle \rightarrow |\tilde{v}(t)\rangle$ is of order $\tilde{\mathcal{O}}(s_C \|H\|_{tot}t)$, where $\|H_{tot}\|=\max(\|C\|_{max} \|D\|_{max}, \|D\|_{max})$ and $\|D\|_{max}=N \sim 1/\epsilon$. For analogue quantum simulation, simulation is with respect to the Hamiltonian $H_{tot}=-(C_1 \otimes D+C_2 \otimes I)$, which might also have potential on trapped ion systems mentioned in Section~\eqref{sec:A}. The initial state is now $|x_0\rangle$ and the cost in the final step of recovering $|x(t)\rangle$ from $|\tilde{v}(t)\rangle$ is of order $\|x(0)\|/\|x(t)\|$, which is identical to Section~\ref{sec:A}. 

This algorithm can be used for general $t$. This also includes preparing the ground state of $C-I$ to fidelity $1-\delta$ for large enough $t \geq t_f$. Here we assume $C-I$ has an orthonormal basis $\{|a_i\rangle_{i=0}^{d}\}$ with corresponding eigenvalues $\{a_i\}_{i=0}^{d}$, which are all negative since $r(C)<1$. Here $|a_0\rangle$ is the true ground state. We can write the initial state as $|x_0\rangle=(1/\|x_0\|)\sum_{i=0}^{d}\sum_{i=0}^d \beta_i|a_i\rangle$. Similarly to Eq.~\eqref{heat-tfinal}, here $t_f \gtrsim (1/(2\Delta_C))\ln((1/(\delta |\beta_0|)))$ where $\Delta_C$ is the spectral gap of $C$. In this case, similarly to Section~\ref{sec:A}, the cost in retrieving $|x(t_f)\rangle$ from $|\tilde{v}(t_f)\rangle$ is $\mathcal{O}(1/|\beta_0|)$.

\section{Quantum  iterative solvers for quantum linear systems of equations}

A linear algebraic system of equations takes the form of $Ay=b$, where $y, b \in \mathbb{R}^d$ and $A$ is a $d \times d$ matrix. It has the analytic solution $y=A^{-1}b$. Solving the quantum linear system of equations is the preparation of the quantum state $|y\rangle \propto A^{-1} |b\rangle$, where the values of the classical vectors are encoded in the amplitudes of the corresponding quantum states. We now use an iterative method to prepare this state. 

We focus on the Jacobi method as an example. This iterative algorithm approximates $A^{-1}b$ as the steady state solution of the evolution in $y$. Let $A=\Lambda+M$, where $\Lambda$ is a diagonal matrix and $M$ consists only of off-diagonal terms. Let the evolution of vector $y_k$ at step $k$ be as
\begin{align}\label{Jacobi}
    y_{k+1}=Gy_k+g, \qquad G=-\Lambda^{-1}M, \quad g=\Lambda^{-1}b.
\end{align}
Since this involves the inverse of $\Lambda$, this method is only applicable when $A$ does not have any zero terms along its diagonal. It can easily be checked that in the steady state, $y_{k+1}=y_{k}$, 
\begin{align}
    y_{k+1}-y_k=\Lambda^{-1}b-(\Lambda^{-1}M+I)y_{k}=0
\end{align}
which implies $\Lambda^{-1}b =(\Lambda^{-1}M+I)y_{k}$ or $A y_{k}=b$. Augmenting $y \rightarrow x=(y, 1)^T$, the linear system of equations problem is equivalent to finding the steady state solution of the dynamical system $dx/dt=(C-I)x, x(0)=x_0$ where 
\begin{align} \label{eq:cmatrixinverse}
    C=\begin{pmatrix}
        G & g \\
        \vect{0}^T & 1
    \end{pmatrix}=\begin{pmatrix}
        -\Lambda^{-1}M & \Lambda^{-1}b \\
        \vect{0}^T & 1
    \end{pmatrix}. 
\end{align}
For the iterative method to converge, we require that the spectral radius $r(C)<1$. For this condition to hold $A$ needs to be diagonally dominant, i.e., each diagonal element of $A$ satisfies $|A_{ii}| \geq \sum_{j \neq i} |A_{ij}|$, which is a sufficient condition for the convergence of the Jacobi method \cite{GvL}.

In this case since $g \neq 0$ in general, then $C \neq C^{\dagger}$ so we use the ground state preparation method in Section~\ref{sec:B} to prepare $|y(t_f)\rangle \approx |y\rangle \propto A^{-1}|b\rangle$ by quantum simulation with respect to the Hamiltonian $H_{tot}$ in Eq.~\eqref{eq:Htot}, i.e., 
\begin{align}
    H_{tot}=-\begin{pmatrix}
        -\Lambda^{-1}M & \Lambda^{-1}b \\
        \vect{0}^T & 1
    \end{pmatrix} \otimes \frac{D-iI}{2}-\begin{pmatrix}
        -(\Lambda^{-1}M)^{\dagger} & \vect{0}^T \\
        (\Lambda^{-1}b)^{\dagger} & 1
    \end{pmatrix} \otimes \frac{D+iI}{2}+I\otimes D.
\end{align}
We can begin with the initial state $|x_0\rangle$ with $x_0=(y_0, 1)^T$. Thus a sparse initial state $x_0$ corresponds to a sparse $x_0$. The cost in the digital quantum simulation step is $\tilde{\mathcal{O}}(s_{H_{tot}} \|C-I\|_{max}/(\epsilon \Delta_C))$. Here $s_{H_{tot}}$ is of the same order as $s_C$ is $\sim \max\{s_{\Lambda^{-1}M}, s_{\Lambda^{-1}b}\} \sim \max\{s_M, s_b\} \sim \max\{s_A, s_b\}$. The max-norm $\|C-I\|_{max} \sim\max\{\|\Lambda^{-1}M\|_{max}, \|\Lambda^{-1}b\|_{max} \}$. To retrieve $|y(t_f)\rangle$ from $|\tilde{w}(t_f)\rangle$ we need a multiplicative cost of $\mathcal{O}(1/|\beta_0|)$ where $|\beta_0|$ is the overlap between the initial state $|x(0)\rangle$ from the true solution $\propto A^{-1}|b\rangle$. As pointed out in Section~\ref{sec:B}, $C_1=(C+C^{\dagger})/2$ being positive definite and diagonalisable in real space are sufficient assumptions for the initial-value problem in Eq.~\eqref{eq:vequation} to be well-posed and no extra boundary conditions are required. 

This algorithm can alternatively be done by the quantum simulation of the hybrid CV-DV unitary  $\exp(iH_{CV-DV}t_{f})$ where $H_{CV-DV}=-C \otimes (\hat{q}-iI)/2-C^{\dagger} \otimes (\hat{q}+iI)/2+I\otimes \hat{q}$.  

\subsection*{Other stationary iterative methods for solving linear system of equations}

Other stationary iterative methods \cite{GvL} for solving the system of linear equations can also be used and our formalism here can in principle also be applied to them. They differ in how the matrix $A$ is split into different matrices. In general, one can write $A=B+N$ where $B$ is assumed to be easily invertible for the method to work. For the Jacobi method, $B=\Lambda$ are the diagonal elements of $A$. There are also other examples, for instance the Richardson method ($B=I/a$, $a \neq 0$), damped Jacobi method ($B=\Lambda/a$, $a \neq 0, 1$), Gauss–Seidel method ($B=\Lambda+L$, $L$ is the strict lower triangular part of $A$), successive over-relaxation (SOR) method  ($B=\Lambda/a+L$, $a \neq 0$) and symmetric successive over-relaxation (SSOR) method ($B=(\Lambda+aL)\Lambda^{-1}(\Lambda+aU)/(a(2-a))$, $a \neq 0, 2$, $U$ is the strict upper triangular part of $A$). However, many of these other methods, like the Gauss-Siedel, require efficient ways to invert non-trivial matrices like $L$, which may not be possible classically in general. A naive approach is to  use the quantum simulation linear system solver described in the preceding section to invert $L$ at each iteration, but it would be more interesting to seek an improved method that does not require another iterations at each iteration step. On the other hand,   these other methods are not significantly faster in the classical case compared to the Jacobi method, and for quantum computing the difference will be even more insignificant so we will not pursue the quantum implementation of these other iterative methods here. 

\section{Quantum power method for approximating the maximum eigenvalue and eigenvector}

Another example of iterative methods in linear algebra are the power  methods, which are used to identify the  largest eigenvalue and its corresponding eigenvector of a matrix \cite{GvL}. Suppose we have a diagonalisable matrix $C$ with real and positive eigenvalues $1 > \lambda_1>\lambda_2>...>\lambda_d$ and corresponding eigenvectors $\{c_i \}_{i=1}^d$, which form an orthonormal basis set. Then any $d$-dimensional vector can be written as $x_0=\sum_{i=1}^d \gamma_i c_i$. Suppose one is interested in finding the largest eigenvalue $\lambda_1$. Then one can  evolve according to $x_k=Cx_{k-1}=C^k x_0$. Then 
\begin{align}
   & x_k=\gamma_1\lambda_1^k c_1+\sum_{i=2}^d \gamma_i \lambda_i^k c_k \\
&=\lambda_1^k\left(\gamma_1c_1+\sum_{i=2}^d \gamma_i\left(\frac{\lambda_i}{\lambda_1}\right)^kc_i\right).
\end{align}
Assume $\lambda_2<\lambda_1$, then for large enough $k$, say $k>K$, the dominant term is $\lambda_1^k \gamma_1 c_1$ and one has  $x_{K} \approx \gamma_1 \lambda_1^K c_1$ and $x_{K+1} \approx \gamma_1 \lambda_1^{K+1}c_1$, both in the direction of the eigenvector $c_1$. Moreover, one can extract the maximum eigenvalue $\lambda_1$ from
\begin{align}
    \lambda_1 \approx \frac{x_K^T x_{K+1}}{x_K^T x_K}=\frac{x_K^T C x_K^T}{x_K^T x_K}=\langle x_K|C|x_K\rangle
\end{align}
where $|x_K\rangle=(1/\|x_K\|)\sum_{i=1}^d (x_K)_i|i\rangle$. Note that this $|x_K\rangle$ also approximates the eigenvector $|c_1\rangle$ of $C$ with corresponding maximum eigenvalue $\lambda_1$. 

We can turn this into a dynamical problem of preparing the state $|x_K\rangle \rightarrow |x(t=t_{max})\rangle \propto \exp((C-I)t_{max})|x_0\rangle$. Since $r(C)<1$, for large enough $t$, $|x(t)\rangle$ will approach the maximum eigenstate. When $C=C^{\dagger}$, we can use the Schr\"odingerisation approach in Section~\ref{sec:A}. Here one needs to perform quantum simulation with respect to the Hamiltonian $C \otimes D$. When $C\neq C^{\dagger}$, we can use the more general approach in Section~\ref{sec:B} to realise the non-unitary evolution $\exp((C-I)t)$ through quantum simulation. Here quantum evolution with respect to the Hamiltonian $H_{tot}=-C \otimes (D-iI)/2-C^{\dagger}\otimes (D+iI)/2+I \otimes D$. As noted in Section~\ref{sec:B}, $C_1=(C+C^{\dagger})/2$ being positive definite and diagonalisable in real space are sufficient assumptions for the initial-value problem in Eq.~\eqref{eq:vequation} to be well-posed and no extra boundary conditions are required. 

Now we must identify $t_{max}$ such that we can approximate $\lambda_1$ to precision $\epsilon$
\begin{align}
    |\langle x(t_{max})|C|x(t_{max})\rangle-\lambda_1|<\epsilon, 
\end{align}
where $|\langle x(t_{max})|C|x(t_{max})\rangle-\lambda_1|=|\text{Tr}(C(|x(t_{max})\rangle \langle x(t_{max})|-|c_1\rangle \langle c_1|))| \leq \sqrt{\text{Tr}(C^{\dagger}C)} \sqrt{2-\mathcal{F}(|x(t_{max})\rangle, |c_1\rangle)}<\epsilon$. This means it is sufficient to ensure that the quantum fidelity  $\mathcal{F}(|x(t_{max})\rangle, |c_1\rangle)=|\langle x(t_{max})|c_1\rangle|^2 \geq 1-\delta$ is large enough, where $\delta=\epsilon^2/(2\text{Tr}(C^{\dagger}C)) \ll 1$. 

The analysis can proceed similarly to the case in ground state estimation of a Hamiltonian, except here we want to ensure large overlap with the maximum eigenstate instead of the minimum eigenstate. Now
\begin{align}
    |x(t)\rangle=\frac{\|x_0\|}{\|x(t)\|}e^{(C-I)t}|x_0\rangle=\frac{1}{\|x(t)\|}\sum_{j=1}^{d}e^{(\lambda_j-1)t}\gamma_j |c_j\rangle
\end{align}
where $\|x(t)\|^2=\sum_{j=1}^{d}|\gamma_j|^2 e^{2(\lambda_j-1) t}=\exp(2(\lambda_1-1)t)(|\gamma_1|^2+|\gamma_2|^2\exp(-2\tilde{\Delta}_Ct)+\tilde{L})$ where $\tilde{\Delta}_C=\lambda_1-\lambda_2$ is the difference between the largest and the second largest eigenvalue of $C$. Here $\tilde{L}=\sum_{j=3}^d |\gamma_k|^2\exp(-2(\lambda_1-\lambda_k))$. Then $\mathcal{F}(|x(t_{max})\rangle, |c_1\rangle)=(|\gamma_1|/\|x(t_{max})\|)^2\exp(2(\lambda_1-1)t_{max}) \geq 1-\delta$ implies
\begin{align}
    t_{max} \gtrsim \frac{1}{2\tilde{\Delta}_C}\ln \left( \frac{2\text{Tr}(C^{\dagger}C)}{\epsilon^2} \left(\frac{1}{|\gamma_1|^2}-1\right)\right)
\end{align}
where we assumed that the eigenstates $|c_1\rangle$, $|c_2\rangle$ dominate the initial state $|x_0\rangle$, i.e., $|\gamma_1|^2 \approx 1-|\gamma_2|^2$. We also assumed $\tilde{L} \ll \delta|\gamma_1|^2/(1-\delta)$. 

Once $|x(t_f)\rangle$ is obtained, we can compute the expectation value $\lambda_1 \approx \langle x(t_f)|C|x(t_f)\rangle$ by estimating $\langle x(t_f)|C_1|x(t_f)\rangle$ and $\langle x(t_f)|C_2|x(t_f)\rangle$ separately, where $C_1=(C+C^{\dagger})/2=C^{\dagger}_1$ and $C_2=(C-C^{\dagger})/(2i)=C^{\dagger}_2$ are identified to be observables. The measurement cost here is the standard $\mathcal{O}(1/\epsilon^2)$ which can also be quadratically boosted, e.g. \cite{rall2020quantum}. 

Here we do not want to retrieve the ground state of $C$ but rather its maximum eigenvector. We can turn to the more general method in Section~\ref{sec:B} and insert $t=t_{max}$. Then the cost for digital quantum simulation and subsequent estimation of the maximum eigenstate $|c_1\rangle$ is $\tilde{\mathcal{O}}(s_C \|C\|_{max}/(\tilde{\Delta}_C|\gamma_1|\epsilon))$. 

This algorithm can alternatively be done by the quantum simulation of the hybrid CV-DV unitary $\exp(iH_{CV-DV}t_{max})$ where $H_{CV-DV}=-C \otimes (\hat{q}-iI)/2-C^{\dagger} \otimes (\hat{q}+iI)/2+I\otimes \hat{q}$. 

We end this section by mentioning that one could use the inverse power and shifted-power methods to find other eigenvalues \cite{GvL}. Like the other iterative methods for linear systems mentioned in preceding section, these methods require one to invert non-trivial matrices like $A$ or $A-\nu I$ (for $\nu$ constants). One could use the naive approach that at each  iteration step,  use the quantum simulation linear system solver described in the preceding section to invert these matrices. However, it would be more interesting to seek an improved method that does not require such a procedure at each iteration step.

 \section{Discussion}

We can first consider a comparison of our quantum Jacobi method to its classical counterpart. The cost in the classical Jacobi method is $\mathcal{O}(s_A d K)$ \cite{GvL} where $K=t_f/\Delta t$ is the total discrete number of iteration steps in the algorithm and $\Delta t$ is the small time step converting the the discrete time step into continuous time $t_f$. For instance, we can take $\Delta t \sim \epsilon$ and $\mathcal{O}(s_A d K) \sim \mathcal{O}(s_A d t_f/\epsilon)$. Here the cost $s_A d$ comes from matrix multiplication at each iteration step. Quantum algorithms do not speed up classical algorithms through a faster $t_f$. Rather, any potential quantum speed-up comes from the more efficient matrix multiplication at each time-step. In the digital quantum simulation protocol, the total cost is $\mathcal{O}(s_A \|\Lambda^{-1}M\|_{max} t_f/(\epsilon |\alpha_0|))$, where $t_f$ remains the same for both the classical and quantum algorithms. Thus in cases where the overlap $|\alpha_0|$ between the initial state and the true solution is high enough, e.g. $\lesssim \mathcal{O}(1/\log(d))$ instead of $\sim 1/d$, then the quantum method can be much more efficient. However, we must emphasise here that our protocol, like HHL \cite{harrow2009quantum}, does not solve $A^{-1}b$, but rather solves the analogous quantum problem of preparing the quantum state $\propto A^{-1}|b\rangle$. This means that if the retrieval of classical solutions from the quantum Jacobi method is also to be exponentially more efficient compared to the classical method, then the number of classical solutions extracted from the final quantum state must also be of order $\sim \log(d)$.

 The most well-known algorithm to solve the quantum linear systems of equations is the HHL algorithm \cite{harrow2009quantum}, which carries a cost $\tilde{\mathcal{O}}(s_A\kappa_A^2/\epsilon)$ where $\kappa_A$ is the condition number for the $d \times d$ matrix $A$. This can be up to exponentially more efficient than classical methods like the conjugate gradient method $\tilde{\mathcal{O}}(d s_A \sqrt{\kappa_A})$ and Gaussian elimination $\mathcal{O}(d^{2.376})$ (excluding costs in initial state preparation) when $A$ is both sparse $s \sim \mathcal{O}(1)$ and well-conditioned $\kappa=\mathcal{O}(\text{poly}(\log d))$. Extensions include more sophisticated methods that provide improvement on $\epsilon$-dependence and $\kappa_A$ to $\mathcal{O}(s_A \kappa_A \text{poly}(\log(s_A \kappa_A/\epsilon)))$ \cite{childs2017quantum}. The effect of the condition number can be further reduced in some cases to $\sqrt{\kappa_A}$ \cite{orsucci2021solving} and preconditioning can also be applied \cite{tong2021fast}.  
 
The quantum Jacobi method differs from these methods above in some important respects. The first is the simplicity of the approach through continuous-time quantum simulation, which does not need to appeal to quantum phase estimation or more sophisticated methods based on approximating a non-unitary operator by a sum of unitaries or to block-encoding \cite{gilyen2019quantum}. Unlike previous methods, our protocol can require just a {\it single copy} of the initial state instead of multiple copies, because it largely replies on unitary evolution, where Grover-like rotations instead of projective measurements may be used in the last step. There is also no longer an explicit dependence on $\kappa_A$ in the Jacobi method, and instead we have a dependence on the spectral gap of $A$ and the overlap between the initial state with the true solution, which provides a complimentary perspective to the canonical methods.

We note that there are also other ground state preparation algorithms, which in principle could also be used for these stationary iterative methods, but to the best of our knowledge Jacobi and power methods have not yet been applied.  These include algorithms based on quantum phase estimation \cite{ge2019faster}, more sophisticated methods relying on block-encoding \cite{lin2020near, an2022theory}, linear combination of unitaries \cite{an2023linear} and quantum adiabatic computation \cite{van2001powerful}. There are also quantum approaches to Krylov subspace iterative methods like the Lanczos method in \cite{kirby2022exact}. These could all provide alternative approaches to the quantum Jacobi method outlined here. 

Next we can consider the comparison of the quantum power method to its classical counterpart, which costs $\mathcal{O}(s_C d K)$ where $K$ is the final discrete number of iterations. Like for the Jacobi method, it can be similarly translated into $\mathcal{O}(s_C d t_f /\epsilon)$ where $t_f$ is the continuous final time. The cost in preparing the quantum eigenstate of $C$ with the quantum power method is $\tilde{\mathcal{O}}(s_C \|C\|_{max}t_f/(|\gamma_1|\epsilon))$. Thus in the case of sparse matrix $C$ and when $|\gamma_1|\sim 1/\log(d)$, this can be up to exponentially more efficient than the classical counterpart. We emphasize that the quantum algorithm for preparing the eigenvector prepares a quantum state, not a classical eigenvector, although a small number e.g. $\log(d)$ number of entries of the classical vector can still be efficiently retrieved. However, the quantum state can be used to efficiently retrieve the maximum eigenvalue, with the total cost $\tilde{\mathcal{O}}(s_C \|C\|_{max}t_f/(\epsilon^2|\gamma_1|))$. 

There are also other quantum algorithms for approximating maximum eigenvectors of density matrices corresponding to large eigenvalues, for instance using quantum principal component analysis \cite{lloyd2014quantum}. This is based on density matrix exponentiation, which requires multiple copes of the density matrix. It also relies on a low-rank approximation, which may be subject to dequantisation results \cite{tang2021quantum}.  More recent work on finding maximum eigenvalues with quantum algorithms is also based on the power method \cite{nghiem2022quantum} but relies on HHL \cite{harrow2009quantum} as a subroutine. 

Our focus here is on quantum simulation that is not aided by classical optimisation algorithms, where the latter is often exploited in numerous hybrid classical-quantum approaches. We will not discuss this literature here. 

While the above analysis has been performed for the discrete case, our scheme is also possible on hybrid CV-DV quantum simulators, which provides an alternative quantum computational framework to solve problems in quantum linear algebra. In this case, the quantum simulation is done with respect to the hybrid CV-DV unitary process $\exp(-i H_{CV-DV}t)$ where $t=t_f, t_{max}$ remains the same as for the fully discrete scheme. Here $\log N$ qubits can be replaced by a single qumode representing the $\eta$ degree of freedom. In the hybrid case, the quantum Fourier transforms is performed on the qumode, which makes it very easy to implement. This evolution can potentially be realised on hybrid systems including photonic circuits, superconducting circuits and trapped ion systems.

\section*{Acknowledgements}
SJ was partially supported by the NSFC grant No.~12031013, the Shanghai Municipal Science and Technology Major Project (2021SHZDZX0102), and the Innovation Program of Shanghai Municipal Education Commission (No. 2021-01-07-00-02-E00087).  NL acknowledges funding from the Science and Technology Program of Shanghai, China (21JC1402900). 

\bibliography{Qiterative}

\begin{thebibliography}{36}%
\makeatletter
\providecommand \@ifxundefined [1]{%
 \@ifx{#1\undefined}
}%
\providecommand \@ifnum [1]{%
 \ifnum #1\expandafter \@firstoftwo
 \else \expandafter \@secondoftwo
 \fi
}%
\providecommand \@ifx [1]{%
 \ifx #1\expandafter \@firstoftwo
 \else \expandafter \@secondoftwo
 \fi
}%
\providecommand \natexlab [1]{#1}%
\providecommand \enquote  [1]{``#1''}%
\providecommand \bibnamefont  [1]{#1}%
\providecommand \bibfnamefont [1]{#1}%
\providecommand \citenamefont [1]{#1}%
\providecommand \href@noop [0]{\@secondoftwo}%
\providecommand \href [0]{\begingroup \@sanitize@url \@href}%
\providecommand \@href[1]{\@@startlink{#1}\@@href}%
\providecommand \@@href[1]{\endgroup#1\@@endlink}%
\providecommand \@sanitize@url [0]{\catcode `\\12\catcode `\$12\catcode
  `\&12\catcode `\#12\catcode `\^12\catcode `\_12\catcode `\%12\relax}%
\providecommand \@@startlink[1]{}%
\providecommand \@@endlink[0]{}%
\providecommand \url  [0]{\begingroup\@sanitize@url \@url }%
\providecommand \@url [1]{\endgroup\@href {#1}{\urlprefix }}%
\providecommand \urlprefix  [0]{URL }%
\providecommand \Eprint [0]{\href }%
\providecommand \doibase [0]{https://doi.org/}%
\providecommand \selectlanguage [0]{\@gobble}%
\providecommand \bibinfo  [0]{\@secondoftwo}%
\providecommand \bibfield  [0]{\@secondoftwo}%
\providecommand \translation [1]{[#1]}%
\providecommand \BibitemOpen [0]{}%
\providecommand \bibitemStop [0]{}%
\providecommand \bibitemNoStop [0]{.\EOS\space}%
\providecommand \EOS [0]{\spacefactor3000\relax}%
\providecommand \BibitemShut  [1]{\csname bibitem#1\endcsname}%
\let\auto@bib@innerbib\@empty
\bibitem [{\citenamefont {Jin}\ \emph {et~al.}(2022)\citenamefont {Jin},
  \citenamefont {Liu},\ and\ \citenamefont {Yu}}]{schr2}%
  \BibitemOpen
  \bibfield  {author} {\bibinfo {author} {\bibfnamefont {S.}~\bibnamefont
  {Jin}}, \bibinfo {author} {\bibfnamefont {N.}~\bibnamefont {Liu}},\ and\
  \bibinfo {author} {\bibfnamefont {Y.}~\bibnamefont {Yu}},\ }\bibfield
  {title} {\bibinfo {title} {Quantum simulation of partial differential
  equations via schr\"odingerisation},\ }\href@noop {} {\bibfield  {journal}
  {\bibinfo  {journal} {arXiv: 2212.13969}\ } (\bibinfo {year}
  {2022})}\BibitemShut {NoStop}%
\bibitem [{\citenamefont {Feynman}(2018)}]{feynman2018simulating}%
  \BibitemOpen
  \bibfield  {author} {\bibinfo {author} {\bibfnamefont {R.~P.}\ \bibnamefont
  {Feynman}},\ }\bibfield  {title} {\bibinfo {title} {Simulating physics with
  computers},\ }in\ \href@noop {} {\emph {\bibinfo {booktitle} {Feynman and
  computation}}}\ (\bibinfo  {publisher} {CRC Press},\ \bibinfo {year} {2018})\
  pp.\ \bibinfo {pages} {133--153}\BibitemShut {NoStop}%
\bibitem [{\citenamefont {Daley}\ \emph {et~al.}(2022)\citenamefont {Daley},
  \citenamefont {Bloch}, \citenamefont {Kokail}, \citenamefont {Flannigan},
  \citenamefont {Pearson}, \citenamefont {Troyer},\ and\ \citenamefont
  {Zoller}}]{daley2022practical}%
  \BibitemOpen
  \bibfield  {author} {\bibinfo {author} {\bibfnamefont {A.~J.}\ \bibnamefont
  {Daley}}, \bibinfo {author} {\bibfnamefont {I.}~\bibnamefont {Bloch}},
  \bibinfo {author} {\bibfnamefont {C.}~\bibnamefont {Kokail}}, \bibinfo
  {author} {\bibfnamefont {S.}~\bibnamefont {Flannigan}}, \bibinfo {author}
  {\bibfnamefont {N.}~\bibnamefont {Pearson}}, \bibinfo {author} {\bibfnamefont
  {M.}~\bibnamefont {Troyer}},\ and\ \bibinfo {author} {\bibfnamefont
  {P.}~\bibnamefont {Zoller}},\ }\bibfield  {title} {\bibinfo {title}
  {Practical quantum advantage in quantum simulation},\ }\href@noop {}
  {\bibfield  {journal} {\bibinfo  {journal} {Nature}\ }\textbf {\bibinfo
  {volume} {607}},\ \bibinfo {pages} {667} (\bibinfo {year}
  {2022})}\BibitemShut {NoStop}%
\bibitem [{\citenamefont {Harrow}\ \emph {et~al.}(2009)\citenamefont {Harrow},
  \citenamefont {Hassidim},\ and\ \citenamefont {Lloyd}}]{harrow2009quantum}%
  \BibitemOpen
  \bibfield  {author} {\bibinfo {author} {\bibfnamefont {A.~W.}\ \bibnamefont
  {Harrow}}, \bibinfo {author} {\bibfnamefont {A.}~\bibnamefont {Hassidim}},\
  and\ \bibinfo {author} {\bibfnamefont {S.}~\bibnamefont {Lloyd}},\ }\bibfield
   {title} {\bibinfo {title} {Quantum algorithm for linear systems of
  equations},\ }\href@noop {} {\bibfield  {journal} {\bibinfo  {journal}
  {Physical review letters}\ }\textbf {\bibinfo {volume} {103}},\ \bibinfo
  {pages} {150502} (\bibinfo {year} {2009})}\BibitemShut {NoStop}%
\bibitem [{\citenamefont {Clader}\ \emph {et~al.}(2013)\citenamefont {Clader},
  \citenamefont {Jacobs},\ and\ \citenamefont
  {Sprouse}}]{clader2013preconditioned}%
  \BibitemOpen
  \bibfield  {author} {\bibinfo {author} {\bibfnamefont {B.~D.}\ \bibnamefont
  {Clader}}, \bibinfo {author} {\bibfnamefont {B.~C.}\ \bibnamefont {Jacobs}},\
  and\ \bibinfo {author} {\bibfnamefont {C.~R.}\ \bibnamefont {Sprouse}},\
  }\bibfield  {title} {\bibinfo {title} {Preconditioned quantum linear system
  algorithm},\ }\href@noop {} {\bibfield  {journal} {\bibinfo  {journal}
  {Physical review letters}\ }\textbf {\bibinfo {volume} {110}},\ \bibinfo
  {pages} {250504} (\bibinfo {year} {2013})}\BibitemShut {NoStop}%
\bibitem [{\citenamefont {Childs}\ \emph {et~al.}(2017)\citenamefont {Childs},
  \citenamefont {Kothari},\ and\ \citenamefont {Somma}}]{childs2017quantum}%
  \BibitemOpen
  \bibfield  {author} {\bibinfo {author} {\bibfnamefont {A.~M.}\ \bibnamefont
  {Childs}}, \bibinfo {author} {\bibfnamefont {R.}~\bibnamefont {Kothari}},\
  and\ \bibinfo {author} {\bibfnamefont {R.~D.}\ \bibnamefont {Somma}},\
  }\bibfield  {title} {\bibinfo {title} {Quantum algorithm for systems of
  linear equations with exponentially improved dependence on precision},\
  }\href@noop {} {\bibfield  {journal} {\bibinfo  {journal} {SIAM Journal on
  Computing}\ }\textbf {\bibinfo {volume} {46}},\ \bibinfo {pages} {1920}
  (\bibinfo {year} {2017})}\BibitemShut {NoStop}%
\bibitem [{\citenamefont {Gily{\'e}n}\ \emph {et~al.}(2019)\citenamefont
  {Gily{\'e}n}, \citenamefont {Su}, \citenamefont {Low},\ and\ \citenamefont
  {Wiebe}}]{gilyen2019quantum}%
  \BibitemOpen
  \bibfield  {author} {\bibinfo {author} {\bibfnamefont {A.}~\bibnamefont
  {Gily{\'e}n}}, \bibinfo {author} {\bibfnamefont {Y.}~\bibnamefont {Su}},
  \bibinfo {author} {\bibfnamefont {G.~H.}\ \bibnamefont {Low}},\ and\ \bibinfo
  {author} {\bibfnamefont {N.}~\bibnamefont {Wiebe}},\ }\bibfield  {title}
  {\bibinfo {title} {Quantum singular value transformation and beyond:
  exponential improvements for quantum matrix arithmetics},\ }in\ \href@noop {}
  {\emph {\bibinfo {booktitle} {Proceedings of the 51st Annual ACM SIGACT
  Symposium on Theory of Computing}}}\ (\bibinfo {year} {2019})\ pp.\ \bibinfo
  {pages} {193--204}\BibitemShut {NoStop}%
\bibitem [{\citenamefont {Suba{\c{s}}{\i}}\ \emph {et~al.}(2019)\citenamefont
  {Suba{\c{s}}{\i}}, \citenamefont {Somma},\ and\ \citenamefont
  {Orsucci}}]{subacsi2019quantum}%
  \BibitemOpen
  \bibfield  {author} {\bibinfo {author} {\bibfnamefont {Y.}~\bibnamefont
  {Suba{\c{s}}{\i}}}, \bibinfo {author} {\bibfnamefont {R.~D.}\ \bibnamefont
  {Somma}},\ and\ \bibinfo {author} {\bibfnamefont {D.}~\bibnamefont
  {Orsucci}},\ }\bibfield  {title} {\bibinfo {title} {Quantum algorithms for
  systems of linear equations inspired by adiabatic quantum computing},\
  }\href@noop {} {\bibfield  {journal} {\bibinfo  {journal} {Physical review
  letters}\ }\textbf {\bibinfo {volume} {122}},\ \bibinfo {pages} {060504}
  (\bibinfo {year} {2019})}\BibitemShut {NoStop}%
\bibitem [{\citenamefont {Lloyd}\ \emph {et~al.}(2014)\citenamefont {Lloyd},
  \citenamefont {Mohseni},\ and\ \citenamefont
  {Rebentrost}}]{lloyd2014quantum}%
  \BibitemOpen
  \bibfield  {author} {\bibinfo {author} {\bibfnamefont {S.}~\bibnamefont
  {Lloyd}}, \bibinfo {author} {\bibfnamefont {M.}~\bibnamefont {Mohseni}},\
  and\ \bibinfo {author} {\bibfnamefont {P.}~\bibnamefont {Rebentrost}},\
  }\bibfield  {title} {\bibinfo {title} {Quantum principal component
  analysis},\ }\href@noop {} {\bibfield  {journal} {\bibinfo  {journal} {Nature
  Physics}\ }\textbf {\bibinfo {volume} {10}},\ \bibinfo {pages} {631}
  (\bibinfo {year} {2014})}\BibitemShut {NoStop}%
\bibitem [{\citenamefont {Nghiem}\ and\ \citenamefont
  {Wei}(2022)}]{nghiem2022quantum}%
  \BibitemOpen
  \bibfield  {author} {\bibinfo {author} {\bibfnamefont {N.~A.}\ \bibnamefont
  {Nghiem}}\ and\ \bibinfo {author} {\bibfnamefont {T.-C.}\ \bibnamefont
  {Wei}},\ }\bibfield  {title} {\bibinfo {title} {Quantum algorithm for
  estimating largest eigenvalues},\ }\href@noop {} {\bibfield  {journal}
  {\bibinfo  {journal} {arXiv preprint arXiv:2211.06179}\ } (\bibinfo {year}
  {2022})}\BibitemShut {NoStop}%
\bibitem [{\citenamefont {Farhi}\ \emph {et~al.}(2000)\citenamefont {Farhi},
  \citenamefont {Goldstone}, \citenamefont {Gutmann},\ and\ \citenamefont
  {Sipser}}]{farhi2000quantum}%
  \BibitemOpen
  \bibfield  {author} {\bibinfo {author} {\bibfnamefont {E.}~\bibnamefont
  {Farhi}}, \bibinfo {author} {\bibfnamefont {J.}~\bibnamefont {Goldstone}},
  \bibinfo {author} {\bibfnamefont {S.}~\bibnamefont {Gutmann}},\ and\ \bibinfo
  {author} {\bibfnamefont {M.}~\bibnamefont {Sipser}},\ }\bibfield  {title}
  {\bibinfo {title} {Quantum computation by adiabatic evolution},\ }\href@noop
  {} {\bibfield  {journal} {\bibinfo  {journal} {arXiv preprint
  quant-ph/0001106}\ } (\bibinfo {year} {2000})}\BibitemShut {NoStop}%
\bibitem [{\citenamefont {Aharonov}\ \emph {et~al.}(2008)\citenamefont
  {Aharonov}, \citenamefont {Van~Dam}, \citenamefont {Kempe}, \citenamefont
  {Landau}, \citenamefont {Lloyd},\ and\ \citenamefont
  {Regev}}]{aharonov2008adiabatic}%
  \BibitemOpen
  \bibfield  {author} {\bibinfo {author} {\bibfnamefont {D.}~\bibnamefont
  {Aharonov}}, \bibinfo {author} {\bibfnamefont {W.}~\bibnamefont {Van~Dam}},
  \bibinfo {author} {\bibfnamefont {J.}~\bibnamefont {Kempe}}, \bibinfo
  {author} {\bibfnamefont {Z.}~\bibnamefont {Landau}}, \bibinfo {author}
  {\bibfnamefont {S.}~\bibnamefont {Lloyd}},\ and\ \bibinfo {author}
  {\bibfnamefont {O.}~\bibnamefont {Regev}},\ }\bibfield  {title} {\bibinfo
  {title} {Adiabatic quantum computation is equivalent to standard quantum
  computation},\ }\href@noop {} {\bibfield  {journal} {\bibinfo  {journal}
  {SIAM review}\ }\textbf {\bibinfo {volume} {50}},\ \bibinfo {pages} {755}
  (\bibinfo {year} {2008})}\BibitemShut {NoStop}%
\bibitem [{\citenamefont {Kulenovic}\ and\ \citenamefont
  {Merino}(2002)}]{KM2002discrete}%
  \BibitemOpen
  \bibfield  {author} {\bibinfo {author} {\bibfnamefont {M.~R.}\ \bibnamefont
  {Kulenovic}}\ and\ \bibinfo {author} {\bibfnamefont {O.}~\bibnamefont
  {Merino}},\ }\href@noop {} {\emph {\bibinfo {title} {Discrete dynamical
  systems and difference equations with Mathematica}}}\ (\bibinfo  {publisher}
  {CRC Press},\ \bibinfo {year} {2002})\BibitemShut {NoStop}%
\bibitem [{\citenamefont {Galor}(2007)}]{galor2007discrete}%
  \BibitemOpen
  \bibfield  {author} {\bibinfo {author} {\bibfnamefont {O.}~\bibnamefont
  {Galor}},\ }\href@noop {} {\emph {\bibinfo {title} {Discrete dynamical
  systems}}}\ (\bibinfo  {publisher} {Springer Science \& Business Media},\
  \bibinfo {year} {2007})\BibitemShut {NoStop}%
\bibitem [{\citenamefont {Andersen}\ \emph {et~al.}(2015)\citenamefont
  {Andersen}, \citenamefont {Neergaard-Nielsen}, \citenamefont {Van~Loock},\
  and\ \citenamefont {Furusawa}}]{andersen2015hybrid}%
  \BibitemOpen
  \bibfield  {author} {\bibinfo {author} {\bibfnamefont {U.~L.}\ \bibnamefont
  {Andersen}}, \bibinfo {author} {\bibfnamefont {J.~S.}\ \bibnamefont
  {Neergaard-Nielsen}}, \bibinfo {author} {\bibfnamefont {P.}~\bibnamefont
  {Van~Loock}},\ and\ \bibinfo {author} {\bibfnamefont {A.}~\bibnamefont
  {Furusawa}},\ }\bibfield  {title} {\bibinfo {title} {Hybrid discrete-and
  continuous-variable quantum information},\ }\href@noop {} {\bibfield
  {journal} {\bibinfo  {journal} {Nature Physics}\ }\textbf {\bibinfo {volume}
  {11}},\ \bibinfo {pages} {713} (\bibinfo {year} {2015})}\BibitemShut
  {NoStop}%
\bibitem [{\citenamefont {Lloyd}(2003)}]{lloyd2003hybrid}%
  \BibitemOpen
  \bibfield  {author} {\bibinfo {author} {\bibfnamefont {S.}~\bibnamefont
  {Lloyd}},\ }\bibfield  {title} {\bibinfo {title} {Hybrid quantum computing},\
  }\href@noop {} {\bibfield  {journal} {\bibinfo  {journal} {Quantum
  information with continuous variables}\ ,\ \bibinfo {pages} {37}} (\bibinfo
  {year} {2003})}\BibitemShut {NoStop}%
\bibitem [{\citenamefont {Liu}\ \emph {et~al.}(2016)\citenamefont {Liu},
  \citenamefont {Thompson}, \citenamefont {Weedbrook}, \citenamefont {Lloyd},
  \citenamefont {Vedral}, \citenamefont {Gu},\ and\ \citenamefont
  {Modi}}]{liu2016power}%
  \BibitemOpen
  \bibfield  {author} {\bibinfo {author} {\bibfnamefont {N.}~\bibnamefont
  {Liu}}, \bibinfo {author} {\bibfnamefont {J.}~\bibnamefont {Thompson}},
  \bibinfo {author} {\bibfnamefont {C.}~\bibnamefont {Weedbrook}}, \bibinfo
  {author} {\bibfnamefont {S.}~\bibnamefont {Lloyd}}, \bibinfo {author}
  {\bibfnamefont {V.}~\bibnamefont {Vedral}}, \bibinfo {author} {\bibfnamefont
  {M.}~\bibnamefont {Gu}},\ and\ \bibinfo {author} {\bibfnamefont
  {K.}~\bibnamefont {Modi}},\ }\bibfield  {title} {\bibinfo {title} {Power of
  one qumode for quantum computation},\ }\href@noop {} {\bibfield  {journal}
  {\bibinfo  {journal} {Physical Review A}\ }\textbf {\bibinfo {volume} {93}},\
  \bibinfo {pages} {052304} (\bibinfo {year} {2016})}\BibitemShut {NoStop}%
\bibitem [{\citenamefont {Golub}\ and\ \citenamefont {Van~Loan}(1996)}]{GvL}%
  \BibitemOpen
  \bibfield  {author} {\bibinfo {author} {\bibfnamefont {G.~H.}\ \bibnamefont
  {Golub}}\ and\ \bibinfo {author} {\bibfnamefont {C.~F.}\ \bibnamefont
  {Van~Loan}},\ }\href@noop {} {\emph {\bibinfo {title} {Matrix
  computations}}},\ \bibinfo {edition} {3rd}\ ed.,\ Johns Hopkins Studies in
  the Mathematical Sciences\ (\bibinfo  {publisher} {Johns Hopkins University
  Press, Baltimore, MD},\ \bibinfo {year} {1996})\ pp.\ \bibinfo {pages}
  {xxx+698}\BibitemShut {NoStop}%
\bibitem [{\citenamefont {Lax}(1973)}]{Lax}%
  \BibitemOpen
  \bibfield  {author} {\bibinfo {author} {\bibfnamefont {P.~D.}\ \bibnamefont
  {Lax}},\ }\href@noop {} {\emph {\bibinfo {title} {Hyperbolic systems of
  conservation laws and the mathematical theory of shock waves}}},\ Conference
  Board of the Mathematical Sciences Regional Conference Series in Applied
  Mathematics, No. 11\ (\bibinfo  {publisher} {Society for Industrial and
  Applied Mathematics, Philadelphia, Pa.},\ \bibinfo {year} {1973})\ pp.\
  \bibinfo {pages} {v+48}\BibitemShut {NoStop}%
\bibitem [{\citenamefont {Braunstein}\ and\ \citenamefont
  {Van~Loock}(2005)}]{braunstein2005quantum}%
  \BibitemOpen
  \bibfield  {author} {\bibinfo {author} {\bibfnamefont {S.~L.}\ \bibnamefont
  {Braunstein}}\ and\ \bibinfo {author} {\bibfnamefont {P.}~\bibnamefont
  {Van~Loock}},\ }\bibfield  {title} {\bibinfo {title} {Quantum information
  with continuous variables},\ }\href@noop {} {\bibfield  {journal} {\bibinfo
  {journal} {Reviews of modern physics}\ }\textbf {\bibinfo {volume} {77}},\
  \bibinfo {pages} {513} (\bibinfo {year} {2005})}\BibitemShut {NoStop}%
\bibitem [{\citenamefont {Adesso}\ \emph {et~al.}(2014)\citenamefont {Adesso},
  \citenamefont {Ragy},\ and\ \citenamefont {Lee}}]{adesso2014continuous}%
  \BibitemOpen
  \bibfield  {author} {\bibinfo {author} {\bibfnamefont {G.}~\bibnamefont
  {Adesso}}, \bibinfo {author} {\bibfnamefont {S.}~\bibnamefont {Ragy}},\ and\
  \bibinfo {author} {\bibfnamefont {A.~R.}\ \bibnamefont {Lee}},\ }\bibfield
  {title} {\bibinfo {title} {Continuous variable quantum information: Gaussian
  states and beyond},\ }\href@noop {} {\bibfield  {journal} {\bibinfo
  {journal} {Open Systems \& Information Dynamics}\ }\textbf {\bibinfo {volume}
  {21}},\ \bibinfo {pages} {1440001} (\bibinfo {year} {2014})}\BibitemShut
  {NoStop}%
\bibitem [{\citenamefont {Van~Loock}\ \emph {et~al.}(2008)\citenamefont
  {Van~Loock}, \citenamefont {Munro}, \citenamefont {Nemoto}, \citenamefont
  {Spiller}, \citenamefont {Ladd}, \citenamefont {Braunstein},\ and\
  \citenamefont {Milburn}}]{van2008hybrid}%
  \BibitemOpen
  \bibfield  {author} {\bibinfo {author} {\bibfnamefont {P.}~\bibnamefont
  {Van~Loock}}, \bibinfo {author} {\bibfnamefont {W.}~\bibnamefont {Munro}},
  \bibinfo {author} {\bibfnamefont {K.}~\bibnamefont {Nemoto}}, \bibinfo
  {author} {\bibfnamefont {T.}~\bibnamefont {Spiller}}, \bibinfo {author}
  {\bibfnamefont {T.}~\bibnamefont {Ladd}}, \bibinfo {author} {\bibfnamefont
  {S.~L.}\ \bibnamefont {Braunstein}},\ and\ \bibinfo {author} {\bibfnamefont
  {G.}~\bibnamefont {Milburn}},\ }\bibfield  {title} {\bibinfo {title} {Hybrid
  quantum computation in quantum optics},\ }\href@noop {} {\bibfield  {journal}
  {\bibinfo  {journal} {Physical Review A}\ }\textbf {\bibinfo {volume} {78}},\
  \bibinfo {pages} {022303} (\bibinfo {year} {2008})}\BibitemShut {NoStop}%
\bibitem [{\citenamefont {Sutherland}\ and\ \citenamefont
  {Srinivas}(2021)}]{sutherland2021universal}%
  \BibitemOpen
  \bibfield  {author} {\bibinfo {author} {\bibfnamefont {R.}~\bibnamefont
  {Sutherland}}\ and\ \bibinfo {author} {\bibfnamefont {R.}~\bibnamefont
  {Srinivas}},\ }\bibfield  {title} {\bibinfo {title} {Universal hybrid quantum
  computing in trapped ions},\ }\href@noop {} {\bibfield  {journal} {\bibinfo
  {journal} {Physical Review A}\ }\textbf {\bibinfo {volume} {104}},\ \bibinfo
  {pages} {032609} (\bibinfo {year} {2021})}\BibitemShut {NoStop}%
\bibitem [{\citenamefont {Elliott}\ \emph {et~al.}(2017)\citenamefont
  {Elliott}, \citenamefont {Gu}, \citenamefont {Thompson},\ and\ \citenamefont
  {Liu}}]{elliott2017continuous}%
  \BibitemOpen
  \bibfield  {author} {\bibinfo {author} {\bibfnamefont {T.~J.}\ \bibnamefont
  {Elliott}}, \bibinfo {author} {\bibfnamefont {M.}~\bibnamefont {Gu}},
  \bibinfo {author} {\bibfnamefont {J.}~\bibnamefont {Thompson}},\ and\
  \bibinfo {author} {\bibfnamefont {N.}~\bibnamefont {Liu}},\ }\bibfield
  {title} {\bibinfo {title} {Continuous variable qumodes as non-destructive
  probes of quantum systems},\ }\href@noop {} {\bibfield  {journal} {\bibinfo
  {journal} {arXiv preprint arXiv:1707.04250}\ } (\bibinfo {year}
  {2017})}\BibitemShut {NoStop}%
\bibitem [{\citenamefont {Zhang}\ \emph {et~al.}(2022)\citenamefont {Zhang},
  \citenamefont {Li},\ and\ \citenamefont {Yuan}}]{zhang2022quantum}%
  \BibitemOpen
  \bibfield  {author} {\bibinfo {author} {\bibfnamefont {X.-M.}\ \bibnamefont
  {Zhang}}, \bibinfo {author} {\bibfnamefont {T.}~\bibnamefont {Li}},\ and\
  \bibinfo {author} {\bibfnamefont {X.}~\bibnamefont {Yuan}},\ }\bibfield
  {title} {\bibinfo {title} {Quantum state preparation with optimal circuit
  depth: Implementations and applications},\ }\href@noop {} {\bibfield
  {journal} {\bibinfo  {journal} {Physical Review Letters}\ }\textbf {\bibinfo
  {volume} {129}},\ \bibinfo {pages} {230504} (\bibinfo {year}
  {2022})}\BibitemShut {NoStop}%
\bibitem [{\citenamefont {Berry}\ \emph {et~al.}(2015)\citenamefont {Berry},
  \citenamefont {Childs},\ and\ \citenamefont
  {Kothari}}]{berry2015hamiltonian}%
  \BibitemOpen
  \bibfield  {author} {\bibinfo {author} {\bibfnamefont {D.~W.}\ \bibnamefont
  {Berry}}, \bibinfo {author} {\bibfnamefont {A.~M.}\ \bibnamefont {Childs}},\
  and\ \bibinfo {author} {\bibfnamefont {R.}~\bibnamefont {Kothari}},\
  }\bibfield  {title} {\bibinfo {title} {Hamiltonian simulation with nearly
  optimal dependence on all parameters},\ }in\ \href@noop {} {\emph {\bibinfo
  {booktitle} {2015 IEEE 56th annual symposium on foundations of computer
  science}}}\ (\bibinfo {organization} {IEEE},\ \bibinfo {year} {2015})\ pp.\
  \bibinfo {pages} {792--809}\BibitemShut {NoStop}%
\bibitem [{\citenamefont {Rall}(2020)}]{rall2020quantum}%
  \BibitemOpen
  \bibfield  {author} {\bibinfo {author} {\bibfnamefont {P.}~\bibnamefont
  {Rall}},\ }\bibfield  {title} {\bibinfo {title} {Quantum algorithms for
  estimating physical quantities using block encodings},\ }\href@noop {}
  {\bibfield  {journal} {\bibinfo  {journal} {Physical Review A}\ }\textbf
  {\bibinfo {volume} {102}},\ \bibinfo {pages} {022408} (\bibinfo {year}
  {2020})}\BibitemShut {NoStop}%
\bibitem [{\citenamefont {Orsucci}\ and\ \citenamefont
  {Dunjko}(2021)}]{orsucci2021solving}%
  \BibitemOpen
  \bibfield  {author} {\bibinfo {author} {\bibfnamefont {D.}~\bibnamefont
  {Orsucci}}\ and\ \bibinfo {author} {\bibfnamefont {V.}~\bibnamefont
  {Dunjko}},\ }\bibfield  {title} {\bibinfo {title} {On solving classes of
  positive-definite quantum linear systems with quadratically improved runtime
  in the condition number},\ }\href@noop {} {\bibfield  {journal} {\bibinfo
  {journal} {Quantum}\ }\textbf {\bibinfo {volume} {5}},\ \bibinfo {pages}
  {573} (\bibinfo {year} {2021})}\BibitemShut {NoStop}%
\bibitem [{\citenamefont {Tong}\ \emph {et~al.}(2021)\citenamefont {Tong},
  \citenamefont {An}, \citenamefont {Wiebe},\ and\ \citenamefont
  {Lin}}]{tong2021fast}%
  \BibitemOpen
  \bibfield  {author} {\bibinfo {author} {\bibfnamefont {Y.}~\bibnamefont
  {Tong}}, \bibinfo {author} {\bibfnamefont {D.}~\bibnamefont {An}}, \bibinfo
  {author} {\bibfnamefont {N.}~\bibnamefont {Wiebe}},\ and\ \bibinfo {author}
  {\bibfnamefont {L.}~\bibnamefont {Lin}},\ }\bibfield  {title} {\bibinfo
  {title} {Fast inversion, preconditioned quantum linear system solvers, fast
  green's-function computation, and fast evaluation of matrix functions},\
  }\href@noop {} {\bibfield  {journal} {\bibinfo  {journal} {Physical Review
  A}\ }\textbf {\bibinfo {volume} {104}},\ \bibinfo {pages} {032422} (\bibinfo
  {year} {2021})}\BibitemShut {NoStop}%
\bibitem [{\citenamefont {Ge}\ \emph {et~al.}(2019)\citenamefont {Ge},
  \citenamefont {Tura},\ and\ \citenamefont {Cirac}}]{ge2019faster}%
  \BibitemOpen
  \bibfield  {author} {\bibinfo {author} {\bibfnamefont {Y.}~\bibnamefont
  {Ge}}, \bibinfo {author} {\bibfnamefont {J.}~\bibnamefont {Tura}},\ and\
  \bibinfo {author} {\bibfnamefont {J.~I.}\ \bibnamefont {Cirac}},\ }\bibfield
  {title} {\bibinfo {title} {Faster ground state preparation and high-precision
  ground energy estimation with fewer qubits},\ }\href@noop {} {\bibfield
  {journal} {\bibinfo  {journal} {Journal of Mathematical Physics}\ }\textbf
  {\bibinfo {volume} {60}},\ \bibinfo {pages} {022202} (\bibinfo {year}
  {2019})}\BibitemShut {NoStop}%
\bibitem [{\citenamefont {Lin}\ and\ \citenamefont {Tong}(2020)}]{lin2020near}%
  \BibitemOpen
  \bibfield  {author} {\bibinfo {author} {\bibfnamefont {L.}~\bibnamefont
  {Lin}}\ and\ \bibinfo {author} {\bibfnamefont {Y.}~\bibnamefont {Tong}},\
  }\bibfield  {title} {\bibinfo {title} {Near-optimal ground state
  preparation},\ }\href@noop {} {\bibfield  {journal} {\bibinfo  {journal}
  {Quantum}\ }\textbf {\bibinfo {volume} {4}},\ \bibinfo {pages} {372}
  (\bibinfo {year} {2020})}\BibitemShut {NoStop}%
\bibitem [{\citenamefont {An}\ \emph {et~al.}(2022)\citenamefont {An},
  \citenamefont {Liu}, \citenamefont {Wang},\ and\ \citenamefont
  {Zhao}}]{an2022theory}%
  \BibitemOpen
  \bibfield  {author} {\bibinfo {author} {\bibfnamefont {D.}~\bibnamefont
  {An}}, \bibinfo {author} {\bibfnamefont {J.-P.}\ \bibnamefont {Liu}},
  \bibinfo {author} {\bibfnamefont {D.}~\bibnamefont {Wang}},\ and\ \bibinfo
  {author} {\bibfnamefont {Q.}~\bibnamefont {Zhao}},\ }\bibfield  {title}
  {\bibinfo {title} {A theory of quantum differential equation solvers:
  limitations and fast-forwarding},\ }\href@noop {} {\bibfield  {journal}
  {\bibinfo  {journal} {arXiv:2211.05246}\ } (\bibinfo {year}
  {2022})}\BibitemShut {NoStop}%
\bibitem [{\citenamefont {An}\ \emph {et~al.}(2023)\citenamefont {An},
  \citenamefont {Liu},\ and\ \citenamefont {Lin}}]{an2023linear}%
  \BibitemOpen
  \bibfield  {author} {\bibinfo {author} {\bibfnamefont {D.}~\bibnamefont
  {An}}, \bibinfo {author} {\bibfnamefont {J.-P.}\ \bibnamefont {Liu}},\ and\
  \bibinfo {author} {\bibfnamefont {L.}~\bibnamefont {Lin}},\ }\bibfield
  {title} {\bibinfo {title} {Linear combination of hamiltonian simulation for
  non-unitary dynamics with optimal state preparation cost},\ }\href@noop {}
  {\bibfield  {journal} {\bibinfo  {journal} {arXiv preprint arXiv:2303.01029}\
  } (\bibinfo {year} {2023})}\BibitemShut {NoStop}%
\bibitem [{\citenamefont {Van~Dam}\ \emph {et~al.}(2001)\citenamefont
  {Van~Dam}, \citenamefont {Mosca},\ and\ \citenamefont
  {Vazirani}}]{van2001powerful}%
  \BibitemOpen
  \bibfield  {author} {\bibinfo {author} {\bibfnamefont {W.}~\bibnamefont
  {Van~Dam}}, \bibinfo {author} {\bibfnamefont {M.}~\bibnamefont {Mosca}},\
  and\ \bibinfo {author} {\bibfnamefont {U.}~\bibnamefont {Vazirani}},\
  }\bibfield  {title} {\bibinfo {title} {How powerful is adiabatic quantum
  computation?},\ }in\ \href@noop {} {\emph {\bibinfo {booktitle} {Proceedings
  42nd IEEE symposium on foundations of computer science}}}\ (\bibinfo
  {organization} {IEEE},\ \bibinfo {year} {2001})\ pp.\ \bibinfo {pages}
  {279--287}\BibitemShut {NoStop}%
\bibitem [{\citenamefont {Kirby}\ \emph {et~al.}(2022)\citenamefont {Kirby},
  \citenamefont {Motta},\ and\ \citenamefont {Mezzacapo}}]{kirby2022exact}%
  \BibitemOpen
  \bibfield  {author} {\bibinfo {author} {\bibfnamefont {W.}~\bibnamefont
  {Kirby}}, \bibinfo {author} {\bibfnamefont {M.}~\bibnamefont {Motta}},\ and\
  \bibinfo {author} {\bibfnamefont {A.}~\bibnamefont {Mezzacapo}},\ }\bibfield
  {title} {\bibinfo {title} {Exact and efficient lanczos method on a quantum
  computer},\ }\href@noop {} {\bibfield  {journal} {\bibinfo  {journal} {arXiv
  preprint arXiv:2208.00567}\ } (\bibinfo {year} {2022})}\BibitemShut {NoStop}%
\bibitem [{\citenamefont {Tang}(2021)}]{tang2021quantum}%
  \BibitemOpen
  \bibfield  {author} {\bibinfo {author} {\bibfnamefont {E.}~\bibnamefont
  {Tang}},\ }\bibfield  {title} {\bibinfo {title} {Quantum principal component
  analysis only achieves an exponential speedup because of its state
  preparation assumptions},\ }\href@noop {} {\bibfield  {journal} {\bibinfo
  {journal} {Physical Review Letters}\ }\textbf {\bibinfo {volume} {127}},\
  \bibinfo {pages} {060503} (\bibinfo {year} {2021})}\BibitemShut {NoStop}%
\end{thebibliography}%
\end{document}